\documentclass[aps,prl,twocolumn,superscriptaddress,nofootinbib,nopacs]{revtex4}
\pdfoutput=1

\usepackage{amsmath,latexsym,amssymb,graphicx,hyperref,color,enumerate}
\usepackage{slashed}
\usepackage{verbatim}
\usepackage{amsfonts}

\hypersetup{colorlinks,citecolor= nicegreen,linkcolor= nicered}
\usepackage{color}
\definecolor{nicered}{rgb}{0.7,0.1,0.1}
\definecolor{nicegreen}{rgb}{0.1,0.5,0.1}

\newcommand\SEC[1]{\smallskip\medskip\noindent{\sf\bfseries #1}}
\newcommand\SUBSEC[1]{\medskip\noindent{\itshape \bfseries #1}}

\input epsf

\begin{document}

\title{Problems with  False Vacua in Supersymmetric Theories }
 
 \author{Borut Bajc} 
\affiliation{ J. Stefan Institute, 1000 Ljubljana, Slovenia}
\affiliation{ Department of Physics, University of Ljubljana, 1000 Ljubljana, Slovenia}

\author{Gia Dvali}
\affiliation{ASC, Department f\"{u}r Physik, LMU, M\"{u}nchen, 
Theresienstr.~37, 80333 M\"{u}nchen, Germany}

 \affiliation{MPI f\"{u}r Physik, F\"{o}hringer Ring 6, 80805 M\"{u}nchen,Germany}

\affiliation{CERN, Theory Division, 1211 Geneva 23, Switzerland}
\affiliation{CCPP, Department of Physics, New York University, Washington Place, New
York, NY 10003, USA}

\author{Goran Senjanovi\'c}
\affiliation{ICTP, 34100 Trieste, Italy}

\date{\today}

\begin{abstract}
\noindent
 It has been suggested recently that in a consistent theory any  Minkowski vacuum must be exactly stable.   As a result, a large class of theories that in ordinary treatment would appear 
 sufficiently long-lived, in reality make no sense. In particular, this applies to supersymmetric models in which global supersymmetry  is broken in a false vacuum.  We show that in any such  theory  the dynamics of supersymmetry breaking cannot be decoupled 
from  the Planck scale physics.   This finding poses an obvious challenge for the idea of low-scale 
metastable (for example gauge) mediation.  
\end{abstract}

\maketitle

 \SEC{Introduction. }
       The stability of our vacuum is a question of life and death, which must be faced
     if there exists a lower energy state. After all, a metastable vacuum would do, if sufficiently long-lived 
   to comply with the physical observations. The question appears an academic one, for even if we
   live in a pre-apocalyptic age, we rest assured the doomsday will not happen tomorrow. The real issue
   is whether or not a physical theory can guarantee us sufficient stability that we observe; if it does not, we must discard it.
  In the  standard view put forward in the classic papers \cite{coleman, cd},  zero energy  vacua can easily be
 metastable with a sufficiently long life-time. However, it was argued recently \cite{gia} that 
any Minkowski  vacuum, which in the standard treatment is regarded as metastable, in reality exhibits instability at an infinite rate, and 
 becomes unphysical. 
 The only consistent  Minkowski vacua are thus exactly stable ones, true even in the presence of lower energy AdS vacua.
  
      This puts severe constraints on any theory with false vacua. The typical situation where one naturally encounters such local vacua 
      takes place in supersymmetric theories. This Letter is devoted to this important question and we show that a large class of seemingly consistent (according to the standard treatment)
  metastable  theories with spontaneously broken supersymmetry,  in fact  make no sense.    

   
       The vacuum that we live in can be taken to be Minkowskian with an 
  extraordinary accuracy, for even if non-vanishing, the cosmological constant is effectively zero at particle physics scales.
  Therefore, in what follows we shall work in the approximation 
  of exact asymptotic Poincare-invariance.  
   
 Although the study of vacuum decay via tunneling has earlier history~\cite{Kobzarev:1974cp}, 
 a systematic study of this issue was given in the classical work by Coleman~\cite{coleman}. 
He focused on a  study of an $O(4)$- symmetric bounce that describes materialization of a  spherically-symmetric bubble of the true vacuum.   A bubble describes a portion of 
   the true vacuum embedded in the false one, with the scalar field interpolating between the 
   true and the false vacua across a layer, which is called the bubble wall.  In the so-called thin-wall approximation the bubble can be characterized by a positive wall tension (energy per unit surface) $T$  and the  negative volume energy of the interior $V_{true}$. 
    The energy of a static bubble then can be approximated as
  \begin{equation}
   M\, = \,  4\pi R^2T  \, - \,  \frac{4}{3} \pi R^3 |V_{true} | \, .    
\label{energy}
\end{equation}   
   Note that by energy conservation a critical bubble that can be materialized in Minkowski space must have zero total energy, which fixes the size of the critical bubble to be 
   $R_c \, = \,  3T/|V_{true}| $.

    However, as shown by Coleman and De Luccia \cite{cd}, in certain cases, when gravity effects are taken into account, the materialization of the bubble never happens. This happens when
   the critical bounce has an infinite action, or equivalently, the critical bubble an
    infinite radius. This effect is referred to as the Coleman-De Luccia (CDL) 
   suppression and in the thin-wall approximation it leads to the bound
   \begin{equation}
     6\pi  \, G_N T^2   \, \geq \,  |V_{true}| \, , 
   \label{cdlbound} 
   \end{equation}
 where   $G_N \, \equiv \, 1/8\pi M_P^2 $ is the Newton's constant  and $M_P$ is the Planck mass.

  This bound can be directly read off from the expression of the bubble energy for finite $M_P$~\cite{weinberg}, 
  which modifies the flat space expression (\ref{energy}) as 
    \begin{eqnarray}
   M &=&   4\pi R^2T \sqrt{1\, + \, \frac{8\pi}{ 3} G_N|V_{true}| R^2}   \, - \,  \frac{4}{3} \pi R^3 |V_{true} |  \nonumber \\
& &   - 8\pi^2G_NT^2R^3  \, . 
\label{energygravity}
\end{eqnarray}   
 The CDL bound is then recovered by demanding that  the energy of the bubble be semi-positive definite,  or equivalently, $R_c$ infinite.

   Since then,  the standard view on the tunneling is  that violation of the bound is not a problem, since although the vacuum becomes  metastable, it can easily be sufficiently long lived.  
   This view was rejected recently in  \cite{gia}, where it was pointed out that any theory that violates 
the CDL bound, and thus allows for an instability of  Minkowski vacuum,  is inconsistent. In other words, the CDL bound is not a stability condition, but rather a 
{\it consistency}  requirement.   In what follows, we shall refer to it as the "Minkowski-safety" condition.

      The argument of \cite{gia} is based on the fact that violation of the bound implies the existence of  negative mass states in the Minkowski vacuum and renders the theory defined on such a vacuum  sensless. 
       This can be seen already from the expression of bubble energy above.      
  Indeed, consider a theory that violates the CDL  bound.  This means that 
  the theory allows a critical bubble of finite size $R_c$. Then, by continuity such a theory will allow 
  bubbles of a bigger size that will have a negative mass. 
  In order to illustrate this point, consider
    a bubble with $R \, \gg \, R_c$ that has zero velocity at some given moment of time
  (but experiences a non-zero acceleration that forces it to expand).   Such a bubble has a  negative 
  mass, as it is obvious from the expressions (\ref{energy}) and (\ref{energygravity}).   Any theory permitting such a bubble is a disaster.  
  A consistent Poincare invariant theory cannot allow states of negative mass.  The  existence of  such states makes the vacuum unstable with an {\it infinite}  rate, because they can be nucleated (for example,  pair-produced)  in combination with positive mass states with an unbounded probability.   The infinity appears due to the divergence in the phase space integration since a negative mass bubble can be pair produced in combination with a positive energy 
 one  with an arbitrary relative momentum. 
 
   The  Euclidean bounces that describe materialization of such bubble pairs in general  do not 
  minimize the action, but this suppression is unimportant due to divergence of the phase space integral. 
 
   This argument tells us that a consistent theory  in a  gravity-decoupling  limit cannot have a zero energy vacuum coexisting with a negative  energy one.  
 Consider, for example, a theory in which the negative energy vacuum survives in 
   $M_P\, \rightarrow \infty$ limit, so that the 
  effect of gravity on the bubble dynamics can be ignored. Then,  according to 
 (\ref{energy}),  
any  bubble larger than the critical  $R \, > \, R_c \, = \,  3T/|V_{true}| $ has a negative mass. 
Such a bubble is not a static solution of the equations of motion,
but bubbles that materialize as a result of quantum tunneling need not be static.  
For example, the critical spherical bubble that carries zero energy is not static either.  
%
 
    In short,  any theory that admits a Minkowski vacuum that violates the CDL bound in inconsistent.  
The purpose of this Letter is to apply this criterion to a class of supersymmetric theories in which supersymmetry breaking takes place in a false vacuum.  
 
 \SEC{Metastable supersymmetry breaking. }
  It is well known that in global supersymmetric theories the energy of the vacuum is semi-positive definite,  being strictly positive for spontaneously broken supersymmetry and zero for unbroken one.   On the other hand, in supergravity even the vacuum with spontaneously broken supersymmetry can have  zero energy.  
 
   This fact enables one to consider an option of spontaneous supersymmetry breaking, in which the dynamics of supersymmetry breaking and its mediation is decoupled from gravity and survives in the limit $M_P \rightarrow \infty$.  In such a case, it is usually assumed that the entire dynamics  of supersymmetry breaking can be addressed in globally supersymmetric limit, with the role of supergravity corrections being reduced to a technical tool for fine tuning the vacuum energy to zero. 
    
     The possibility of tuning the vacuum energy to zero allows to consider scenarios 
     in which spontaneous dynamical supersymmetry breaking in the global limit takes place in a false vacuum \cite{meta}.  This scenario goes under the name of metastable supersymmetry breaking and has certain simplifications from the point of view of model building 
     (e.g., avoiding constraints from the Witten index \cite{wittenindex}).   Naively, it seems that, since the vacuum energy has to be anyway tuned  by supergravity, in the global limit 
the non-supersymmetric vacuum could be just a local minimum. We shall show that 
 this intuition is false, and that any dynamical supersymmetry breaking  vacuum that survives in the 
 gravity-decoupling limit must be the {\it true} one. 
 

    In what follows,  we start first with a general argument and then illustrate it on a concrete example.

  \SUBSEC{The general argument.} 
    Let $X$ be a chiral superfield that spontaneously breaks supersymmetry through an expectation 
    value of its $F$-term, $F_X$.   We shall assume that  in global-supersymmetry limit supersymmetry is broken in a classically-stable false  vacuum, where $F_X \, = \, \Lambda^2$, and  $X = X_0$, and that there  also exists a true supersymmetry-preserving vacuum in which $F_X =0$.  Since for our considerations what matters is the difference of scales, without loss of generality we can assume that  
    in the true vacuum $X=0$.
    
     Our assumptions will be that  the mass scales $\Lambda$ and $X_0$ that determine particle masses and VEVs  are small compared to  the Planck mass $M_P$ and  that masses of scalars in these two minima  are much larger than the gravity-mediated soft mass, 
   \begin{equation}
      m_X^2 \, \gg \,  \frac{\Lambda^4}{M_P^2} \, .
     \label{condition}
  \end{equation}
  This is a necessary condition for  sub-dominance of gravity mediation. 
 Usually, this follows from the condition $X_0 \, \ll \, M_P$, but we shall spell it out separately for clarity. 
  This condition implies that the back-reaction through the effects of gravity on the expectation values in the vacua of interest is small. 
  
   Let us now switch on supergravity.  The scalar potential becomes (we ignore possible $D$-terms)  
  \begin{equation}
    V(X) \, = \, e^{\frac{K}{M_P^2}} \left ( (K_X^X)^{-1}  F_XF^X \, - \, 3 |W|^2/M_P^2 \right) \, ,
    \label{pot}
    \end{equation} 
where $K$ is the K\"ahler function,  and upper (lower) index $X$ stands for 
derivative with respect to  $X$ ($X^*$) respectively.   In terms of superpotential 
$W$ and the K\"ahler  metric the $F$-terms are given as 
 \begin{equation}
   F_X = \,  (F^X)^*\,  = \,   D_XW  \,  \equiv \,   W_X \,+\, K_X  W /M_P^2 \, .
    \label{fterm}
    \end{equation} 
    In order to have a supersymmetry-breaking Minkowski vacuum,   the negative term in (\ref{pot}) 
    must  cancel the positive one.    But since all  VEVs are sub-Planckian, we have 
      $K_X/M_P \,  \sim \, X_0/M_P \, \ll \, 1$. So the tuning of the vacuum energy implies 
   (up to corrections of order $X_0/M_P$)      
   $W_X \, = \Lambda^2\,=\, \sqrt{3}  W  /M_P$ , 
     which is accomplished by introducing the following constant in the superpotential, 
\begin{equation}
W_0 \, = \, \Lambda^2M_P/\sqrt{3}\,.
\label{constant}
\end{equation} 
But now, the supersymmetry-preserving minimum 
  appears to have a negative energy equal to 
      \begin{equation}
     V_{true}  \,   =  \, - \Lambda^4 \, .
    \label{true}
    \end{equation} 
 The two minima are separated by a barrier that is set by a scale $\Lambda^4$, and the 
 length of the barrier is at most $X_0$. This situation badly violates the Coleman-De Luccia bound, since the tension of the domain wall is at most   $ T \sim \Lambda^4/m_X$.   This is because the 
 energy density in the wall is $\Lambda^4$ and the thickness is  $m_X^{-1}$. 
  So we have  
  \begin{equation}
  T^2 /M_P^2 \, \ll \, |V_{true}| \, , 
  \label{CDL}
  \end{equation}
   and the Minkowski safety constraint cannot be satisfied.  
%
%
\SUBSEC{ An explicit example.}  
Consider a simple case that reproduces the essential features of the model 
  of \cite{ddgr} of metastable  supersymmetry-breaking.
   Take two  chiral superfields $X$ and $\Phi$,  with the superpotential 
      \begin{equation}
        W \,   =  \,  \frac{1}{2} X\Phi^2 \, - \, X\, \Lambda^2 \, .
     \label{super}
    \end{equation} 
 Let us analyse the global supersymmetry case first.  For a minimal K\"ahler  $K \, = \, |X|^2 \, + \,  |\Phi|^2$, the theory has a global supersymmetry-respecting vacuum with $X = 0$ and $\Phi^2 = 2\Lambda^2$,  and a plateau  for  $\Phi =0$ and $|X| \,  > \, \Lambda$, along which $F_X = \Lambda^2$.  
  In the supersymmetry-preserving vacuum the masses of the particles are $m_{X}^2 \,  = \, 2\Lambda^2$ and 
  $m_{\Phi}^2\, = \, 2\Lambda^2$. Whereas along the supersymmetry-violating plateau, the masses of the 
  real and imaginary components of the complex $\Phi$-scalar are  split 
  $m_{\Phi \pm }^2 \, =  |X|^2 \,  \pm \, \Lambda^2$, while the mass of the $X$-modulus is zero. 
  Following \cite{ddgr}, we shall now create a classically-stable local minimum on the plateau 
  at some  $X = X_0$. For this we have to modify the K\"ahler metric,  in such a way, that 
  the function $(K_X^X)^{-1}$ has a minimum at some $X\, = \, X_0$. 
   In \cite{ddgr} this was achieved by the perturbative renormalization of the K\"ahler (i.e. wave-function
   renormalization)  due to the loops of chiral and gauge multiplets.  The effect of this renormalization can be summed up 
   in the following form: 
        \begin{equation}
     (K_X^X)^{-1} \,    =  \,  1 \,  -  \, \epsilon_1 \, \log{\left (\frac{ |X|}{ \Lambda} \right )}  \, + \, \epsilon_2 \, \log^2{ \left (\frac{|X|}{\Lambda} \right )} \, , 
     \label{krun}
    \end{equation} 
  where, $\epsilon_1$ and $\epsilon_2$ are the one- and the two-loop factors respectively.  So 
  approximately  $\epsilon_2 \, \sim   \, \epsilon_1^2$.   In order to have a metastable minimum, 
  the parameters must be chosen in such a way that both coefficients are positive. 
 In \cite{ddgr} this was achieved by the interplay of the gauge and  Yukawa couplings, 
 in the spirit of Witten's  inverted hierarchy idea \cite{witten}. 
    In the global supersymmetric limit,  the stable minimum then develops  at:
 \begin{equation}
  |X| \, =  \,  X_0  \, = \,  \Lambda \, \exp\left({\frac{\epsilon_1}{2 \epsilon_2}}\right)  \, . 
     \label{mini1}
    \end{equation} 
 The mass of the $X$-modulus in this minimum is,  $m_X^2 \, = \, \epsilon_2 \Lambda^4/X_0^2$, whereas the phase of $X$ is a Goldstone boson of a spontaneously broken $R$-symmetry.   The masses of the 
  real and imaginary components of the complex $\Phi$-scalar are  split, 
  $m_{\Phi \pm }^2 \, =  \, X_0^2 \,  \pm \, \Lambda^2$. 
 
       Let us now take into the account the supergravity corrections. 
  In order for these corrections to have a negligible effect on supersymmetry breaking (and in particular, for gauge-mediation to dominate) we have  to demand that the condition  (\ref{condition})  be satisfied. Our general argument at the end of the 
 previous subsection then applies and
%
    the vacuum energy difference ends up
violating the CDL condition. The theory is inconsistent, as argued in \cite{gia}. 

 It is easy to convince ourselves that the inconsistency we are discovering is not an artifact 
 of a particular form of the K\"ahler metric, but rather is fundamental for any form that creates 
 a metastable vacuum in globally supersymmetric limit, which is not separated from the 
 supersymmetric vacuum  by a Planck distance. 
    Replacing the K\"ahler in (\ref{krun}) by any smooth functional dependence that creates 
   a minimum around $X_0$ gives the same result. This is obvious from the fact that 
   an expansion around the point $X_0$ recovers  all the features of the above example. 
    We leave it up to the reader to try different forms. 
    

 The only way out would be to either violate the condition (\ref{condition})  or  
eliminate the globally-supersymmetric  vacuum altogether.    

\SEC{Summary and Outlook.} 
We reach a surprisingly strong conclusion: in false-vacuum supersymmetry breaking, the contribution 
of gravity cannot be ignored.   
This finding imposes a severe constraint on supersymmetry-breaking  
model building, and in particular creates an obvious obstacle for phenomenologically acceptable low scale
gauge-mediated metastable supersymmetry breaking. 
%
 
 This is not necessarily a no-go theorem,  and with some imagination  
 one could attempt to build a working model of this sort. For example, in the above case 
 one may give up on perturbative calculable  corrections to K\"ahler 
 and instead rely on  Planck scale corrections for creating a stable minimum at $X_0\, \sim \, M_P$. 
  One may even arrange  a low-scale gauge mediation by coupling $X$ to a messenger superfield 
 $Q$, with a bare mass $M_Q$, 
 \begin{equation}
   (X\, - \, M_Q) \bar{Q}Q \, , 
 \label{messenger}
 \end{equation}
  and fine tuning the latter, in such a way that  in a stable minimum  $X_0\, - \, M_Q \,  \ll \, M_P$.
  But such a construction defeats the cause, since it gives up both the
  calculability in terms of low energy parameters as well as the naturalness for which low scale mediation is invoked  to start with. 
 
  Putting aside such options,  supersymmetry-breaking must either take place in a true vacuum (such as in the O'Rafearteigh model) or  be of  high-scale. 
   The latter possibility happens naturally  in perturbative theories without singlets, where supersymmetry is broken in metastable vacuum by $F$-terms of grand unified scale fields.  The already large unification scale tends to be automatically increased in minimal examples \cite{susygut}, and such theories could survive our constraint. 
   
   Our findings are still in full agreement with the general proof that, in supergravity Minkowski 
 vacua, the energy is semi-positive definite \cite{sugra}. The starting condition of these 
 theorems is an existence of a well-defined Hilbert space and of a conserved supercharge.  
   What we are showing is that, although in broken supersymmetry naively one can have negative energy states, these  not only destroy the Hilbert space,  but the vacuum itself.  
   In other words,  Minkowski vacuum cannot allow the negative energy states irrespectively of 
   supersymmetry.

     In short, one can say that the observation of \cite{gia} makes the requirement of  energy-positivity in any sensible Minkowski vacuum obvious, since it shows that any such vacuum admitting negative mass objects  is not a metastable state, but is rather infinitely short lived, and  thus has no meaningful description in terms of a  Hilbert space.  Such are the vacua which violate
CDL bounds, and must be excluded. 

One last comment, likely to be obvious to the reader. Strictly speaking, our vacuum is not exactly Minkowski,  but this only matters in the case when the size of the critical bubble becomes comparable to the Hubble horizon.  In the case under study, when one worries about 
ending up in the universe with a large cosmological constant (even on present particle physics scales), this clearly plays a negligible role and does not affect our results.
 
\SEC{Acknowledgements:}
We are grateful to Alex Vilenkin for useful discussions regarding the issue of Minkowski-safety. We thank Talal Chowdhury, Alejandra Melfo and Miha Nemev\v sek for their interest in our work and a careful reading of the manuscript.
B.B. thanks LMU and Humboldt Foundation for hospitality during the final stage of  this work. 
The work of B.B. has been supported by the Slovenian Research Agency. 
The work of G.D. was supported in part by Humboldt Foundation under Alexander von Humboldt Professorship,  by European Commission  under 
the ERC advanced grant 226371,   by TRR 33 \textquotedblleft The Dark
Universe\textquotedblright\   and  by the NSF grant PHY-0758032.  The work of G.S. has been supported in part by the EU grant UNILHC-Grant Agreement
PITN-GA-2009-237920.
G.S. also acknowledges the hospitality of the BIAS Institute during the final
stages of this work.

\end{document}